\documentclass[lettersize,journal]{IEEEtran}
\usepackage{amsmath,amsfonts}
\usepackage{algorithmic}
\usepackage{algorithm}
\usepackage{array}
\usepackage[caption=false,font=normalsize,labelfont=sf,textfont=sf]{subfig}
\usepackage{textcomp}
\usepackage{stfloats}
\usepackage{url}
\usepackage{verbatim}
\usepackage{graphicx}
\usepackage{subfig}
\usepackage{multirow}
\usepackage{booktabs}
\usepackage{siunitx}
\usepackage{tabularx}
\usepackage{enumitem}

\usepackage[natbib, bibencoding=utf8, citestyle=numeric, bibstyle=ieee, maxbibnames=999, maxcitenames=2, mincitenames=1, sortcites]{biblatex}
\bibliography{bibliography}

\usepackage[capitalize]{cleveref}

\begin{document}

\title{CoughPhase-CLR: Designing an acoustics-informed foundation model for coughing sound classification}

\author{Marius Moldovan, Anton Batliner, Thomas M. Berghaus, \\Bj\"orn W. Schuller~\IEEEmembership{Fellow,~IEEE}, Andreas Triantafyllopoulos~\IEEEmembership{Member,~IEEE}
\thanks{
M.M., A.T., A.B., and B.S. are with CHI -- the Chair of Health Informatics at the TUM University Hospital, Munich, Germany.
}
\thanks{
A.T., A.B., and B.S. are also with MCML -- the Munich Center for Machine Learning and MDSI -- the Munich Data Science Institute, Munich, Germany.
}
\thanks{
T.B. is with the University Hospital Augsburg at the University of Augsburg, Augsburg, Germany and with the Medical Faculty, Ludwig Maximilians University of Munich, Munich, Germany.
}
\thanks{
B.S. is also with GLAM -- the Group on Language, Audio, \& Music at Imperial College London, London, United Kingdom.
}
\thanks{
This work has received funding from the DFG Reinhard-Kosseleck Grant No.\ (AUDI0NOMOUS).
Part of these results were published as M.M.'s B.Sc. thesis at the Technical University of Munich.
}
\thanks{
Correspondence: andreas.triantafyllopoulos@tum.de
}
}


\maketitle

\begin{abstract}
In this work, we introduce CoughPhase-CLR, a self-supervised learning framework designed to leverage the physiological phases of a cough for robust representation learning. Unlike generic contrastive frameworks, CoughPhase-CLR constructs positive pairs based on these specific acoustic phases. We pre-trained our model on approximately 40 hours of public cough audio and evaluated it across five downstream tasks, including COVID-19 detection, chronic obstructive pulmonary disease (COPD) state classification, and smoker status prediction. Our results demonstrate that cough-specific pre-training consistently outperforms standard random-cropping techniques when training on cough recordings. Additionally, we benchmarked a diverse set of state-of-the-art models on COPD state classification, highlighting the difficulty of this task. The best-performing models, pretrained on either general audio or respiratory sounds, achieved a UAR of 57\%, failing to outperform the state-of-the-art performance of 84\% UAR achieved using speech analysis. 

\end{abstract}


\section{Introduction}
\IEEEPARstart{R}{espiratory} sounds are an important source of information when it comes to diagnosing and monitoring respiratory disorders~\citep{Triantafyllopoulos23-H4H}.
Coughing, wheezing, or abnormal breathing are often the result of some underlying disorder. 
For instance, coughing is a reflective reaction that can be used to expel mucus, germs or irritants, while wheezing results from an obstruction in the airways.
For this reason, the detection of such `abnormal' sounds has long been proposed as a non-invasive biomarker of pathology~\citep{Triantafyllopoulos23-H4H, OPERA, tuberculosisPahar2021}.

However, the presence of these sounds is, by itself, not a definitive indication of pathology.
Both coughing and wheezing can be normal reactions to the accumulation of allergens or dehydration.
More importantly, even in cases where abnormal respiratory sounds are caused by some pathology, it is not directly evident which of the many plausible diagnoses is directly responsible.
Asthma, respiratory infections, chronic obstructive pulmonary disease (COPD), COVID-19 and lung cancers are some of the many potential disorders which can cause these abnormal sounds \citep{waveletAnalysisCough, CoughCOPDvsCovid, OPERA, Triantafyllopoulos23-H4H}.
Consequently, their `mere' detection is only the first step towards diagnosis, and can, at best, serve as a first warning sign that calls for a closer investigation.

Diagnosing the underlying pathology based on respiratory sounds alone is a much harder task.
As we discuss in \cref{sec:related}, previous works have shown promising results in the binary setup of distinguishing between `normal' and `pathological' sounds.
Despite its clinical utility as an early-warning system, this approach still lags behind our ultimate goal -- that of an advanced diagnostic system which can pinpoint the underlying pathology among the many other, alternative diagnoses.
In addition, the current state-of-the-art leverages foundation models~\citep{OPERA, HeAR}, which are pretrained on large datasets of respiratory (and other) sounds using standard self-supervised learning (SSL) techniques.
While promising, these `naive' approaches depend on vast quantities of pretraining data which cannot be easily assembled in a healthcare setup due to data privacy regulations and the relative scarcity of data, especially when considering rare diseases.

In the present contribution, we focus on the most widely-used type of respiratory sound -- coughing.
Coughing is widely-collected as part of structured, in-the-lab data collection paradigms, while numerous works target its in-the-wild detection using automatic algorithms.
Our key contribution is \emph{CoughPhase-CLR}, a novel pretraining framework inspired from the physiology of coughing.
Using this framework, we develop a new foundation model for the classification of coughing sounds, which outperforms the previous state-of-the-art, especially in the low-data regime.
Importantly, we also evaluate performance on a much more challenging, but also more realistic task than previous work: distinguishing between exacerbation and recompensation of COPD. 

The remainder of this work is structured as follows:
\cref{sec:related} presents previous work on coughing analysis.
\cref{sec:methodology} outlines our framework, while \cref{sec:results} presents our key results.
This is followed by a discussion in \cref{sec:discussion} and a presentation of our concluding remarks in \cref{sec:conclusion}.
The code to reproduce our training can be found in \url{https://github.com/CHI-TUM/CoughPhase-CLR} and our best trained model is available for use in \url{https://huggingface.co/CHI-TUM/CoughPhase-CLR}.

\section{Related Work}
\label{sec:related}

This section reviews related work.
We begin with a discussion of the acoustic structure of cough signals, followed by their use in classifying pathology, and ending with a discussion of recently-introduced foundation models for cough signal analysis.

\subsection{Acoustic Structure of Cough Signals}
\begin{figure*}[t!]
    \centering
    \subfloat[Cough event with three phases.]{\includegraphics[width=.48\textwidth]{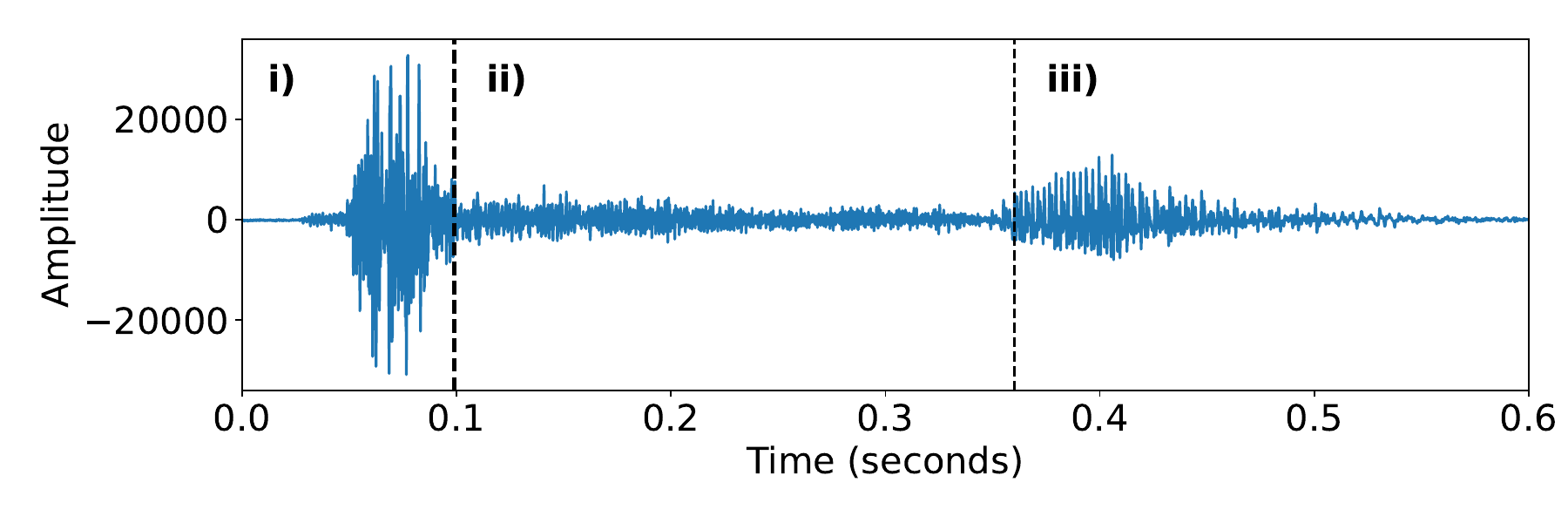}}~%
    \subfloat[Cough event with two phases.]{\includegraphics[width=.48\textwidth]{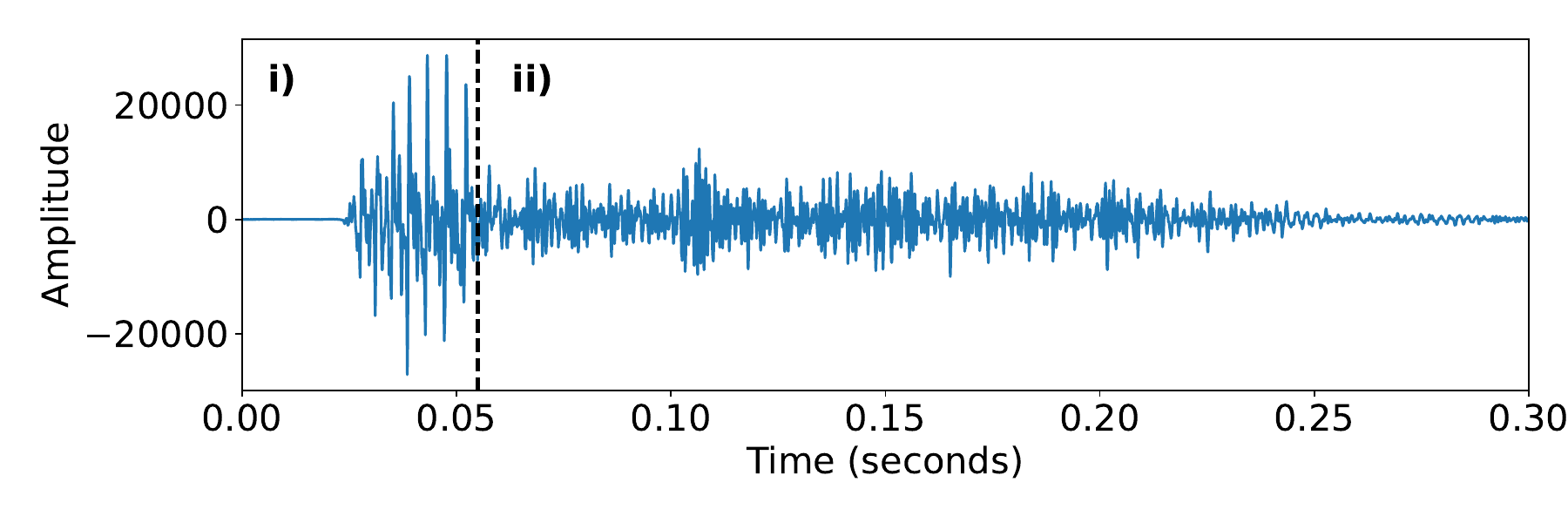}}
    \caption{Example of two coughs divided into the acoustic cough phases: i) explosive, ii) intermediate, and iii) voiced. Adapted from \citet{CoughPhases}.}
    \label{fig:cough_phases}
\end{figure*}
Coughing is a fundamental protective reflex, essential for clearing secretions and foreign particles from the airways. It can be initiated as a voluntary act or as a spontaneous reflex and lasts, on average, 350\,ms \citep{Olia2000CoughStats}. A cough can generally be divided into three phases, as shown in \cref{fig:cough_phases}:
\begin{enumerate}
    \item The \textit{explosive phase} is characterized by an explosive burst with a very sharp, high-amplitude increase in the sound energy envelope. This is caused by the sudden release of pressure as the glottis opens \citep{CoughPhases, Kelsall2008CoughPhases}.
    \item The \textit{intermediate phase} follows this initial burst and is characterized by a more sustained, high-frequency ``noisy'' sound generated by the sustained turbulent airflow through the airways. The energy during this phase is typically lower than the initial burst and gradually decays as airflow decreases \citep{Serrurier2022IntermediatePhase,Hall2020IntermediatePhase}.
    \item The \textit{voiced phase} is the final phase and is not present in all coughs. It includes a pitch frequency induced by the partial closure of the vocal cords, introducing a periodic, tonal quality to the signal \citep{Serrurier2022IntermediatePhase}.
\end{enumerate}

Coughing sounds are also indicative of an underlying pathology.
For instance, coughs from COPD patients typically have a longer duration, a later-occurring peak at lower frequencies, a higher absolute peak, and more energy in the lower frequency ranges compared to healthy controls \citep{waveletAnalysisCough}. Furthermore, the frequency at which the maximum sound energy occurs is significantly lower in COPD patients compared to COVID-19 patients \citep{CoughCOPDvsCovid}.

Additionally, cough sound characteristics show high variability across different ages, biological sex, sputum status, and the presence of an underlying pathology.
Features can often vary more within an individual than between different patients with the same disease or even between patients of different diseases \citep{CoughImportanceCOPD}.
Contributing to this cough variety within individuals are the facts that cough intensity is highest in the morning due to the mucus buildup during the night, and that there is a difference in characteristics between forced coughs and spontaneous ones \citep{CoughPhaseImportanceCovid}.

\subsection{Cough Signal Classification}

Recent work on cough sound classification has rapidly expanded, especially following the COVID-19 pandemic and the increased availability of large-scale open datasets such as \emph{COUGHVID} \citep{Coughvid}, \emph{UK COVID-19} \citep{UK_COVID19}, and \emph{Coswara} \citep{Coswara}. While coughs have emerged as an informative modality for disease classification, there have been mixed results on their effectiveness compared with other vocal modalities like breathing or speech.

Initial experiments on the \emph{Coswara} dataset \citep{Coswara} revealed that coughs (AUC 0.73 to 0.79) were significantly less informative than vowel sounds (AUC 0.93 to 0.94) when performing gender classification, suggesting that coughs are more complex signals to analyze.
However, further research on the same dataset by \citet{CoughSpeechBreathCoswaraAndOthers} showed that, using their method, coughs were better suited to distinguish COVID-19 from healthy controls, with an AUC of 0.98 compared to 0.94 for breathing and 0.92 for speech, concluding that coughs carried the strongest COVID-19 signature.

Cough classification also extends beyond COVID-19.
In a study on tuberculosis, \citet{tuberculosisPahar2021} collected 1,356 cough events from 51 symptomatic patients in a real-world noisy environment.
For the classification of the disease, they achieved an AUC of 0.94 by using logistic regression on MFCC features, outperforming more complex models like CNNs on this small dataset.
Their research also found that increasing the number of MFCC features boosted performance, whereas segmenting the cough audio into different phases offered no benefit, indicating that ``the acoustic information in all phases of a cough is equally important for the purposes of TB classification'' \citep{tuberculosisPahar2021}.

Recent research has also explored the use of transformer architectures for cough classification.
\citet{CoughClassificationTransformer} compared a Swin-Transformer to a shallow 2D-CNN for classifying wet vs dry coughs from spectrograms using the \emph{COUGHVID} \citep{Coughvid} dataset.
The transformer achieved an accuracy of 88.4\%, outperforming the CNN model, which achieved an accuracy of 77.0\%.
This study highlighted the use of augmentations, as time masking alone improved accuracy by 4 percentage points with the full augmentation achieving a 12 percentage point increase.

\subsection{Foundation Models for Coughing}

\citet{OPERA} introduced \textit{OPERA}, an open pre-training and benchmarking system for respiratory acoustics models. The dataset used for pre-training contains a total of around 136,000 respiratory audio samples, totaling over 400 hours. This includes data from different audio signals, including cough, breathing, exhalation, and lung sounds, with the cough recordings coming from the \emph{COUGHVID} \citep{Coughvid}, \emph{UK COVID-19} \citep{UK_COVID19}, and \emph{COVID-19 Sounds} \citep{Cambridge_COVID-19} datasets. They pre-trained a total of three models:
\begin{enumerate}
    \item \emph{OPERA-CE}: \emph{EfficientNet-B0}, trained using a contrastive approach.
    \item \emph{OPERA-CT}: A transformer model, also trained using a contrastive approach.
    \item \emph{OPERA-GT}: A transformer model, trained using a generative approach.
\end{enumerate}
The proposed models were evaluated on 19 downstream tasks and compared against four baselines: the \emph{eGeMAPS} feature set \citep{eGeMAPS}, and three existing audio foundation models (\emph{VGGish} \citep{VGGish}, \emph{AudioMAE} \citep{AudioMAE}, and \emph{CLAP} \citep{CLAP}). For the evaluation, they used a total of 10 labelled datasets, with six datasets being entirely new and unseen during the pre-training phase. They used a linear-probing protocol, training only a linear classifier on top of the frozen model to better assess the quality of the learnt representations. Their results showed improved performance compared with the baseline models in 16 out of the 19 tasks. Moreover, they observed that the model pre-trained with a contrastive objective generally outperformed the generative model on classification tasks, but performed worse on regression tasks. Saliency maps revealed that the contrastive models tend to focus on local, discriminative features in the spectrograms, while generative models learn to capture more global patterns. Additionally, they concluded that while the transformer-based models generally outperformed the CNN-based model, the more lightweight CNN-based model still delivered satisfactory results.

However, a subsequent study by \citet{DiverseDatasetBetterThanOPERA} challenged the idea of pre-training exclusively on respiratory audio. They used seven of the classification tasks from the \emph{OPERA} benchmark and evaluated the performance of 21 audio foundation models using the same methodology as in the \emph{OPERA} paper. The evaluated models included respiratory-specific models (\emph{OPERA-CT}, \emph{OPERA-GT}), speech-focused self-supervised models (\emph{wav2vec2.0}, \emph{HuBERT}), and general audio models pre-trained on \emph{AudioSet} (\emph{ATST-FRAME}, \emph{M2D}). They found that models pre-trained on the larger and more diverse audio dataset \emph{AudioSet} are generally more effective for respiratory tasks than models pre-trained exclusively on smaller respiratory audio datasets. In total, eight of the nineteen models pre-trained on general audio outperformed \emph{OPERA-CT}. The best performance was achieved by further pre-training M2D on a combination of \emph{AudioSet} and additional respiratory audio data, yielding an average AUROC of 0.814 compared to \emph{OPERA-CT}'s 0.733.

In a similar fashion, \citet{COPDCoughSAMOPERA} further investigated the use of different pre-trained models for cough classification. They benchmarked various models pre-trained on either \emph{ImageNet} (e.\,g., a \emph{ResNet} and small \emph{ViT}s), general audio data (e.\,g., \emph{PaSST-S} and \emph{EAT-large}), or respiratory sounds (\emph{OPERA} models). Their proposed model \emph{Cough Search} is a variant of \emph{EAT-large} that was further pre-trained on large-scale cough datasets using both self-supervised and supervised learning techniques in a teacher-student framework. They evaluated the models on three cough datasets: \emph{UK COVID-19} \citep{UK_COVID19}, \emph{COUGHVID} \citep{Coughvid}, and the private \emph{LucaCough} dataset. To preprocess the audio, they extracted single coughs from the recordings by detecting peaks and extracting a window of 5 frames around them. Their experiments showed that models pre-trained on audio data consistently outperform models pre-trained on \emph{ImageNet}. Cough Search achieved the best performance across all three downstream tasks, demonstrating that additional pre-training on cough data can further improve performance.
They also found that using the sharpness-aware minimizer (SAM) \citep{SAM} alongside the Adam optimizer consistently improved performance for all tested models.

\section{Methodology}
\label{sec:methodology}
This section presents our proposed approach as well as our experimental setup.

\subsection{CoughPhase-CLR Framework}
\label{sec:CoughPhase-CLR}

\begin{figure*}[t!]
    \centering
    \includegraphics[width=0.95\textwidth]{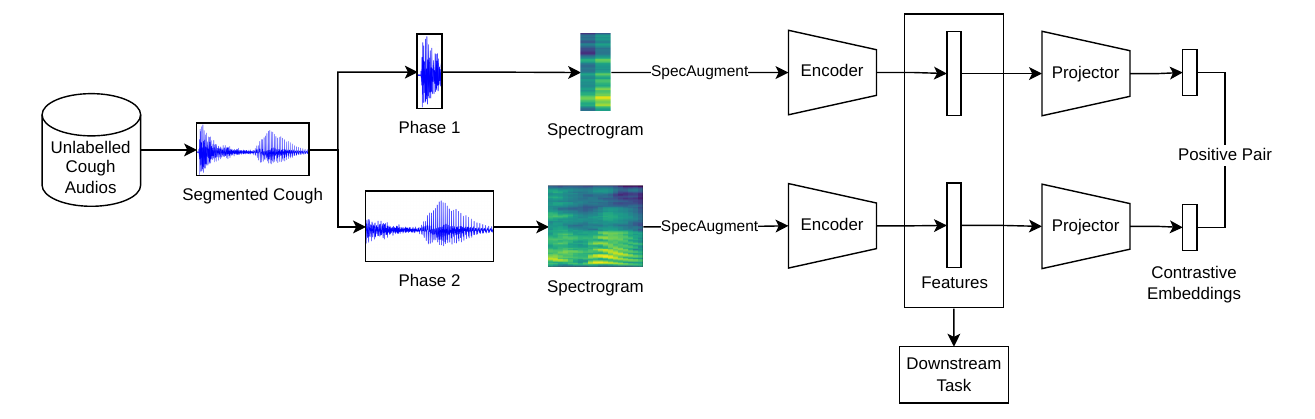}
    \caption{Pre-training framework of \emph{CoughPhase-CLR}.}
    \label{fig:custom_model}
\end{figure*}

In this subsection, we present our \emph{CoughPhase-CLR} -- a contrastively pre-trained model specifically designed to learn robust representations from cough sounds.
Our key motivation is as follows: by training the model to recognize the relationship between the different phases of a cough event, we aim to extract embeddings that capture the most discriminative features of coughs.
The choice of contrastive pre-training is further motivated by prior research demonstrating that the contrastive pre-trained \emph{OPERA} models showed better performance than the generatively pre-trained \emph{OPERA-GT} on classification tasks \citep{OPERA, COPDCoughSAMOPERA}.
The overall framework is illustrated in \cref{fig:custom_model}.

Conceptually, our framework is similar to \emph{OPERA-CE}, with the main difference being the construction of positive and negative pairs during contrastive learning.
Our contrastive task was defined to learn an embedding space in which different phases of the same cough are close to each other, while phases from different coughs are far apart.
To achieve this, we constructed positive pairs from distinct segments of a single cough event.
Although a cough physiologically consists of three phases, we define two segments for our task by treating the initial explosive part as the first segment and combining the subsequent intermediate and voiced phases into a second segment. This was done for two reasons. Firstly, the third, voiced phase is not present in every cough, which makes a two-phase split the more consistent choice for the pre-training.
Secondly, in initial experiments, it was not possible to consistently distinguish between the last two phases in an automated way.
All other samples within a training batch then served as the negative samples.

Our model uses an \emph{EfficientNet-B0} backbone \citep{EfficientNet} as the encoder, selected for its lightweight architecture which enables faster training compared to larger transformer-based models.
To further improve robustness, we applied SpecAugment to all of the spectrograms, using the parameters listed in \cref{tab:SpecAugment_parameters}.
During pre-training, we used a batch size of 256, matching the batch size used in \emph{OPERA-CE}. A batch size of 1024 was also explored in initial experiments, as research suggests that a higher batch size generally achieves better performance for contrastive learning \citep{ChenSimCLR}.
However, a larger batch size showed worse performance in initial tests on the downstream tasks and was later abandoned.

\subsection{Pre-training Datasets}
\label{sec:pretrain-data}
\begin{table*}[t!]
    \centering
    
    \caption{Overview of datasets used for pre-training. (SR: sampling rate; Duration: mean [95\% quantile range])}
    \label{tab:pre-train_Data}
    \small
    \begin{tabular}{lc|c|c|c|c}
        \hline
        Dataset Name                  & \#Recordings & \#Coughs & Modality & SR     & Duration of coughs (s) \\
        \hline
        \emph{UK COVID-19} \citep{UK_COVID19} & 19533        & 47,220   & Cough    & 48\,kHz & 0.60 [0.43 - 0.96]     \\
        \emph{COUGHVID} \citep{Coughvid}      & 7179         & 85,799   & Cough    & 48\,kHz & 0.61 [0.42 - 1.02]     \\
        \hline
    \end{tabular}
\end{table*}
To pre-train \emph{CoughPhase-CLR}, we used the two openly available cough datasets \emph{COUGHVID} \citep{Coughvid} and \emph{UK COVID-19} \citep{UK_COVID19}, with relevant information on the datasets found in \cref{tab:pre-train_Data}.
\emph{COVID-19 Sounds} \citep{Cambridge_COVID-19} was not available to us when conducting our study due to licensing.
We combined both datasets and reserved 10\% for validation and preprocessed the data in the following ways before training:

First, all audio recordings were resampled to a sampling rate of 16\,kHz and the individual cough events were isolated from the longer audio recordings. For the extraction of coughs, the method proposed in \citet{Coughvid} was used. This method detects the beginning and end of coughs based on their energy content, using a hysteresis comparator to mark the onset when the energy exceeds a high threshold and the offset when it falls below a lower one.
An additional 0.2 seconds were added to both ends to capture the whole cough more consistently.

Next, each cough was divided into two distinct segments. While the first segment consists of the original explosive phase, we combined the intermediate and voiced phases into a single second segment. We split the audio into two segments by detecting the biggest peak in the first 60\% of the audio and then splitting it 20\,ms after the peak, to capture the full explosive phase.
Finally, Mel spectrograms were created separately for each cough segment. We used a window size of 1024, a hop length of 512, and 64 Mel frequency bins.

\subsection{Evaluation Datasets}
We evaluated the downstream performance of \emph{CoughPhase-CLR} on three datasets taken from the \emph{OPERA} benchmark \citep{OPERA}, as well as a private COPD dataset.

\noindent
\textbf{UK COVID-19:} The \emph{UK COVID-19} dataset is a public dataset designed to train and evaluate machine learning models for classifying SARS-CoV-2 infection status and respiratory symptoms using audio \citep{UK_COVID19}. The data collection was conducted by the UK Health Security Agency (UKHSA) and the Alan Turing Institute, with participants being recruited via the national Test and Trace program and REACT-1 survey in England between March 2021 and March 2022. The dataset includes audio recordings of forced coughs, exhalations, and speech, along with metadata such as participant demographics, self-reported symptoms, respiratory conditions, and SARS-CoV-2 PCR test results. In total, it includes recordings from 67,842 individuals, of whom 23,514 had positive and 44,328 negative SARS-CoV-2 PCR tests at the time of recording. The recordings were submitted through the ``Speak up to help beat coronavirus'' digital survey, with 97\% of participants providing a linked PCR test result. Out of all participants, 45.62\% reported respiratory symptoms, 11.30\% reported having asthma, and 27.20\% had linked influenza PCR test results. All submitted recordings went through a filtering pipeline, with 5,157 recordings being removed due to missing PCR tests, self-inconsistent symptom responses, or various other reasons.

\noindent
\textbf{COUGHVID:} The \emph{COUGHVID} dataset is a large, publicly available crowdsourced dataset of cough recordings collected through a web application between April and December of 2020 \citep{Coughvid}. The interface followed a simple one-click design and captured up to 10 seconds of audio, with an optional metadata questionnaire including age, respiratory conditions, geographic location, and self-reported COVID-19 status. In total, the dataset contains over 25,000 cough recordings, of which 1,155 were self-reported as COVID-19 positive. Additionally, 2,800 samples were expert-labelled by four experienced physicians, making it one of the largest clinically annotated cough datasets available.
The dataset also provides open-source code for cough segmentation.
Labels include cough type, audible symptoms (wheezing, dyspnea, etc.), and diagnostic impressions (e.\,g., COVID-19, asthma).
Geolocated samples further confirm that most COVID-19 cases came from high-prevalence countries at the time.

\noindent
\textbf{Coswara:} The \emph{Coswara} dataset is an open-access, crowdsourced collection of respiratory sound recordings, created to aid in the remote screening of SARS-CoV-2 \citep{Coswara}.  Through a web application, recordings were collected between April 2020 and February 2022 from a total of 2,635 individuals, with 1,819 being SARS-CoV-2 negative, 674 positive, and 142 recovered at the time of recording. It includes a total of nine types of respiratory sounds, including various kinds of coughs, breathing sounds, sustained vowels, and counting. Additionally, the dataset includes metadata such as gender, smoking status, and an overall COVID status, which was either self-reported or determined by a PCR test. All 23,700 recordings were manually scored for their audio quality, with 87.6\% being excellent, 11.7\% moderate, and 9.7\% poor. All recordings were fixed to a maximum duration of 30 seconds, with the cough recordings having a median duration of 5 seconds. Furthermore, 90\% of the recordings were sampled at 48\,kHz.

\noindent
\textbf{COPD-DE:} The \emph{COPD-DE} dataset is a private collection of audio recordings from patients admitted to the University Hospital of Augsburg with exacerbations of COPD~\citep{TriantafyllopoulosVowels, COPDSpeech2Triantafyllopoulos}. Data collection occurred between October 2020 and February 2023, with recordings taking place before treatment (exacerbation state) and upon discharge (stable state), with an average of $9.1(\pm 4.3)$ days between the two recordings. The study received approval from the local ethics committee on 24 June 2020 (BKF 2020-34). The audio recordings took place at the bedside with a portable Zoom H5 recorder and a Sony lapel mic ECM-144 under controlled settings and quiet surroundings. Four modalities were recorded: forced coughing, sustained vowels, breathing, and reading. In total, 50 patients (male: 26, female: 24) were recorded, of whom 48 (male: 25, female: 23) provided cough recordings at both admission and discharge. From these 48 patients, a total of 166 cough recordings were collected, with 89 being recorded during exacerbation states and 77 during stable states. The average length of the cough recordings is $3.12(\pm 2.4)$ seconds, with a minimum of 0.4 seconds and a maximum of 12.4 seconds.
Alongside the recordings, additional data such as scores on the Borg and the CAT scales \citep{BORG, CAT}, pack-years of smoking, and results of pulmonary function testing done at discharge were collected.
At the time of the first recording, the average Borg score was $4.4(\pm 2.8)$ and the CAT score was $27.9(\pm 5.5)$. On discharge, both scores were lower, with $2.0(\pm 1.8)$ for Borg and $22.2(\pm 6.0)$ for CAT. Steroids were given to 98\% and additional antibiotics to 58\% of all patients.

\subsection{Baselines}
To provide a thorough evaluation, we selected a diverse set of models with different architectures and pre-training strategies for our baselines.
As a traditional machine learning baseline model, we extracted the \emph{eGeMAPS} feature set \citep{eGeMAPS} and used these features for classification with a linear classifier.
As further baseline models, we trained two CNN-based models, namely \emph{CNN14} \citep{PANNs} and \emph{EfficientNet-B0} \citep{EfficientNet}, both from a random initialization (random seed fixed in code for reproducibility). 
Additionally, we incorporated the following pre-trained models:
\begin{itemize}[align=left]
    \item[\emph{ImageNet Pre-training:}] The models pre-trained on \emph{ImageNet} consist of \emph{EfficientNet-B0} (\emph{ImageNet}) and VGG-16-BN \citep{VGG}.

    \item[\emph{General Audio Pre-training:}] For the models pre-trained on general audio data, we chose \emph{CNN14} (\emph{AudioSet}) \citep{PANNs}, as well as \emph{wav2vec2.0} \citep{wav2vec2.0} and \emph{HuBERT} \citep{HuBERT}, which were both pre-trained solely on speech from English audiobooks.

    \item[\emph{Respiratory Audio Pre-training:}] We use two sets of models pre-trained on respiratory audio: the \emph{OPERA} models (\emph{OPERA-CE}, \emph{OPERA-CT}, and \emph{OPERA-GT}) \citep{OPERA}, which were trained on a large dataset of respiratory audio, and the HeAR model \citep{HeAR}, which was trained on respiratory events from YouTube videos.
\end{itemize}

Note that all original \emph{OPERA} variants were trained on a lot more data than our \emph{CoughPhase-CLR}.
To provide a more fair comparison, we retrained \emph{OPERA-CE} on the same pre-training data as our model (see \cref{sec:pretrain-data}).
This resulted in a new model initialization, which we denote as \emph{OPERA-CE-Cough}.

\subsection{Preprocessing}
\label{sec:preprocessing}

\begin{description}
     \item[\emph{OPERA models, CoughPhase-CLR:}] To prepare the audio recordings for the \emph{OPERA} models and for our \emph{CoughPhase-CLR}, we convert them into Mel spectrograms through the following process. First, the length of the raw audio was unified to 4.09 seconds. For files longer than this time, we discarded the end of the audio, while for files shorter than this time, we repeated the audio until the desired length was reached. We then extracted Mel spectrograms with 64 frequency bins using a Hann window of size 1024 and a hop length of 512. This resulted in spectrograms of size $128 \times 64$ pixels, which were then standardized to have zero mean and unit standard deviation. To prevent data leakage, the normalization parameters are always calculated on the training dataset and then applied to the training, development, and test sets. This is the process used by the \emph{OPERA} framework \citep{OPERA}.

    \item[\emph{EfficientNet-B0:}] For the \emph{EfficientNet-B0} model, we extracted the same spectrograms as for the \emph{OPERA} model and subsequently resized them to the model's required input resolution of $224 \times 224$ pixels using bilinear interpolation. For \emph{EfficientNet-B0} (\emph{ImageNet}), we did not apply the standardization calculated on our dataset, but instead normalized the spectrograms using the \emph{ImageNet} mean and standard deviation.

    \item[\emph{CNN14:}] For the \emph{CNN14} model, we instead extracted log-Mel spectrograms with 64 frequency bins using a Hann window of size 512 and a hop length of 160. We then took a random crop over 300 time frames, which results in a spectrogram of size $300 \times 64$ pixels and normalized the spectrograms using instance-level normalization, which subtracts the mean and divides by the standard deviation of each spectrogram. We chose this to match the preprocessing of the original paper \citep{PANNs}.

    \item[\emph{wav2vec2.0, HuBERT, HeAR:}] These models do not take spectrograms as input, but rather require the raw audio waveforms. As a result, we used random 3-second segments from each audio file as inputs, with shorter files being padded with silence.
\end{description}

\subsection{Downstream Training}
Each model was trained for a total of 100 epochs using early stopping, with the best model being selected based on the highest performance on the development set.
Our general approach was to fine-tune the entire model, however, for \emph{HeAR} the weights of the encoder were not updated and only the classifier was trained.
To determine the optimal hyperparameters, we conducted a grid search, with the search space being outlined in \cref{tab:grid_search_hyperparameters}. For the learning rate, we tested values of $10^{-2}$, $10^{-3}$, and $10^{-4}$. We tested batch sizes of 32 and 64, however, memory constraints limited the \emph{VGG-16-BN}, \emph{OPERA-CT}, and \emph{OPERA-GT} models to a batch size of 32. We evaluated two different optimizers. The first one was AdamW with a weight decay of $10^{-2}$, which is one of the most commonly used optimizers for training deep learning models. As a second optimizer, we chose SAM-SGD \citep{SAM} with a neighborhood radius of $\rho=0.05$ and SGD as a base optimizer, as it showed strong performance in an initial test on the \emph{COPD-DE} dataset using \emph{OPERA-CE}.
To address the class imbalance noted in \cref{tab:folds_class_distribution}, Balanced Cross-Entropy was employed as the loss function, where the loss for each class is weighted according to its inverse frequency in the training set.
To further improve model robustness and generalization, we applied data augmentation to the training spectrograms. We decided to use SpecAugment \citep{SpecAugment}, which is a widely used data augmentation technique for spectrograms. The parameters used for SpecAugment are shown in \cref{tab:SpecAugment_parameters}.

\begin{table}[h]
    \centering
    \caption{Overview of hyperparameters used in grid search.}
    \label{tab:grid_search_hyperparameters}
    \begin{tabularx}{1\columnwidth}{l | p{6cm} }
        \toprule
        Hyperparameter & Value                                                                                                                                                                                \\
        \midrule
        Method         & \emph{eGeMAPS}, \emph{EfficientNet-B0} (scratch), \emph{EfficientNet-B0} (\emph{ImageNet}), \emph{CNN14} (scratch), \emph{CNN14} (\emph{AudioSet}), \emph{VGG-16-BN}, \emph{OPERA-CE}, \emph{OPERA-CT}, \emph{OPERA-GT}, \emph{wav2vec2.0}, \emph{HuBERT}, \emph{HeAR}, \emph{CoughPhase-CLR} \\
        Optimizer      & AdamW (weight decay of $10^{-2}$), SAM (base optimizer SGD and $\rho = 0.05$)                                                                                                        \\
        Learning Rates & $10^{-2}$, $10^{-3}$, $10^{-4}$                                                                                                                                                      \\
        Batch Size     & 32, 64 (only 32 for \emph{VGG-16-BN}, \emph{OPERA-CT} and \emph{OPERA-GT})                                                                                                                                \\
        Epochs         & 100                                                                                                                                                                                  \\
        Seed           & 1,2,3                                                                                                                                                                                \\
        \bottomrule
    \end{tabularx}

\end{table}

\begin{table}[h]
    \centering
    \caption{SpecAugment parameters.}
    \label{tab:SpecAugment_parameters}
    \begin{tabular}{l|c c}
        \toprule
        \textbf{Parameter}                & \textbf{Downstream} & \textbf{Pre-training} \\
        \midrule
        Time Warp Window ($W$)            & 10                  & 10                    \\
        Frequency Mask Width ($F$)        & 16                  & 16                    \\
        Time Mask Width ($T$)             & 20                  & 10                    \\
        Number of Frequency Masks ($m_F$) & 1                   & 1                     \\
        Number of Time Masks ($m_T$)      & 1                   & 1                     \\
        \bottomrule
    \end{tabular}
\end{table}

\subsection{Evaluation}
\label{sec:pretrain_evaluation}
\begin{table}[t]
    \caption{Overview of the datasets and tasks used for the downstream evaluation of the pre-trained models. The table details the classification task, the total number of samples, and the class distribution for each dataset. }
        \label{tab:pretrain_evaluation}
    \centering
    \begin{tabular}{l l l r c}
        \hline
        Dataset                   & ID & Task                  & \#Samples & Class Distribution \\ \hline
        \multirow{2}{*}{\emph{COUGHVID}} & T1 & COVID / Non-COVID     & 6175      & 547 / 5628         \\
                                  & T2 & Female / Male         & 7263      & 2468 / 4795        \\ \hline
        \emph{COPD-DE}                      & T3 & Exacerbation / Stable & 166       & 89 / 77            \\ \hline
        \multirow{2}{*}{\emph{Coswara}}  & T4 & Smoker / Non-smoker   & 948       & 201 / 747          \\
                                  & T5 & Female / Male         & 2496      & 759 / 1737         \\ \hline
    \end{tabular}
\end{table}
We benchmarked \emph{CoughPhase-CLR} against all baselines across five downstream tasks using the \emph{COPD-DE}, \emph{Coswara}, and \emph{COUGHVID} datasets (see \cref{tab:pretrain_evaluation}). All downstream tasks used in the evaluation can be found in \cref{tab:pretrain_evaluation}, with the tasks on the \emph{Coswara} and \emph{COUGHVID} datasets being the same ones as used in \emph{OPERA} \citep{OPERA}.

To ensure a rigorous evaluation, we compared our model to both the original \emph{OPERA-CE} model and \emph{OPERA-CE-Cough}, which we created by re-pretraining \emph{OPERA-CE} on the same cough datasets as chosen for our proposed method to gain a direct comparison of the contrastive tasks used. It should be noted though, that while for \emph{CoughPhase-CLR} the individual coughs were extracted, \emph{OPERA-CE-Cough} was pre-trained on the entire audio recordings, which tend to include multiple coughs with silence between them. An overview of all pre-trained models used in the evaluation can be found in \cref{tab:pretrain_evaluation_methods}. Additionally, we also evaluated the use of the \emph{eGeMAPS} \citep{eGeMAPS} features as a baseline.
To ensure a more accurate comparison of the extracted audio representations, the models were not further fine-tuned. Instead, the embeddings were extracted, and a linear layer was trained on top of them.
We used the same evaluation protocol as in \citet{OPERA}, training the model on five different seeds and reporting the AUROC (area under the receiver operating characteristic).

\begin{table}[h]
    \centering
    \caption{Data and contrastive task used for the evaluated pre-trained models.}
    \label{tab:pretrain_evaluation_methods}
    \begin{tabular}{l | l l}
        \toprule
        Method                 & Pre-training data  & Contrastive task \\ \midrule
        \emph{OPERA-CE}  \cite{OPERA} & Respiratory events & Random crop      \\
        \emph{OPERA-CE-Cough}         & Cough              & Random crop      \\
        \emph{CoughPhase-CLR} (ours)  & Cough              & Phase-aware crop \\
        \bottomrule
    \end{tabular}
\end{table}

Finally, we compared models on COPD exacerbation classification using our private \emph{COPD-DE} data \citep{TriantafyllopoulosSpeech}.
Because our dataset is fairly small, with only 166 samples, we implemented 4-fold cross-validation
to obtain a more robust performance estimate from our limited \emph{COPD-DE} dataset.
As the dataset contains multiple recordings for every patient, we implemented a patient-level splitting of the dataset, which means that all recordings from one patient are always assigned to the same data split.
This ensures that the model has to generalize well to new patients and cannot simply learn patient-specific characteristics.
Participants were split according to the date of recruitment, with the first 20 participants forming the test set of the first fold, the next 20 forming the test set of the second fold, and so on.
The resulting class distributions of each fold can be found in \cref{tab:folds_class_distribution}.
We measured the performance of the models using the unweighted average recall (UAR),
as in the original work \citep{TriantafyllopoulosSpeech, TriantafyllopoulosVowels}.
This metric is chosen to account for class imbalance, and is also known as ``balanced accuracy''.

\begin{table}[h]
    \caption{Distribution of samples per class across the cross-validation folds. }
    \label{tab:folds_class_distribution}
    \centering
    \begin{tabular}{c|c c}
        \toprule
        Fold & Exacerbation Count & Stable Count \\
        \midrule
        1    & 17                 & 16           \\
        2    & 16                 & 18           \\
        3    & 17                 & 23           \\
        4    & 39                 & 20           \\
        \bottomrule
    \end{tabular}
\end{table}

\section{Results}
\label{sec:results}

\subsection{Evaluation of CoughPhase-CLR}
\label{sec:evaluation_cough_phase_clr}
\begin{table}[t]
    \caption{Comparison of pre-trained models. Results are given in AUROC. The best performance for each task is in \textbf{bold}, and the second-best is \underline{underlined}. The evaluated tasks are T1 (COVID / Non-COVID, \emph{COUGHVID}), T2 (Female / Male, \emph{COUGHVID}), T3 (Exacerbation / Stable, \emph{COPD-DE}), T4 (Smoker / Non-smoker, \emph{Coswara}), and T5 (Female / Male, \emph{Coswara}).}    \label{tab:downstream_tasks_results}
    \centering
    \begin{tabular}{lccccc}
        \toprule
        \textbf{Method}       & \textbf{T1}      & \textbf{T2}      & \textbf{T3}      & \textbf{T4}      & \textbf{T5}      \\
        \midrule
        \emph{eGeMAPS}               & \underline{0.54} & \underline{0.68} & 0.52             & 0.53             & 0.75             \\ \hline
        \emph{OPERA-CE} \citep{OPERA} & \textbf{0.57}    & \textbf{0.72}    & \textbf{0.61}    & \underline{0.67} & \textbf{0.80}    \\
        \emph{OPERA-CE-Cough}        & 0.53             & \underline{0.68} & 0.58             & 0.62             & 0.71             \\
        \emph{CoughPhase-CLR} (ours) & \textbf{0.57}    & 0.66             & \underline{0.60} & \textbf{0.68}    & \underline{0.76} \\
        \bottomrule
    \end{tabular}

\end{table}

This section evaluates our proposed pre-trained model \emph{CoughPhase-CLR} on the five downstream tasks detailed in \cref{sec:pretrain_evaluation}. We compared its performance against \emph{OPERA-CE}, \emph{OPERA-CE-Cough} and the \emph{eGeMAPS} feature set, with all results being summarized in \cref{tab:downstream_tasks_results}.
Among the three pre-trained models, \emph{OPERA-CE} shows the best performance, with the highest AUROC in four out of five tasks. While our proposed model, \emph{CoughPhase-CLR}, did not consistently surpass \emph{OPERA-CE}, it was able to achieve the highest performance in two out of five tasks, equaling the score from \emph{OPERA-CE} in T1.
\emph{CoughPhase-CLR} did, however, show superior performance to \emph{OPERA-CE-Cough}, outperforming it in four out of five tasks.
This result suggests that our phase-aware contrastive pre-training strategy is more effective for learning from cough audio compared to the random cropping approach used by \emph{OPERA-CE} when both are solely trained on coughs.
\emph{OPERA-CE-Cough} shows the worst performance out of the pre-trained models, achieving the lowest UAR in 2/5 tasks, even falling behind the traditional \emph{eGeMAPS} baseline in most tasks.

\subsection{Scaling Performance of CoughPhase-CLR}
While \emph{CoughPhase-CLR} outperformed \emph{OPERA-CE-Cough} in our overall evaluation, it still lagged behind \emph{OPERA-CE} which was trained on more data.
This raises the question of which form of pre-training scales better with more pre-training data.
To assess this, we pre-trained \emph{CoughPhase-CLR} and \emph{OPERA-CE-Cough} on smaller subsets of the pre-training data and reevaluated their performance on the five downstream tasks.
Specifically, we created smaller subsets of our pre-training dataset, containing 20\%, 40\%, 60\%, and 80\% of the total dataset size, while keeping the ratio between the two datasets the same.

\begin{figure*}[t]
    \centering
    \subfloat[\emph{COUGHVID} COVID-19]{\includegraphics[width=.24\textwidth]{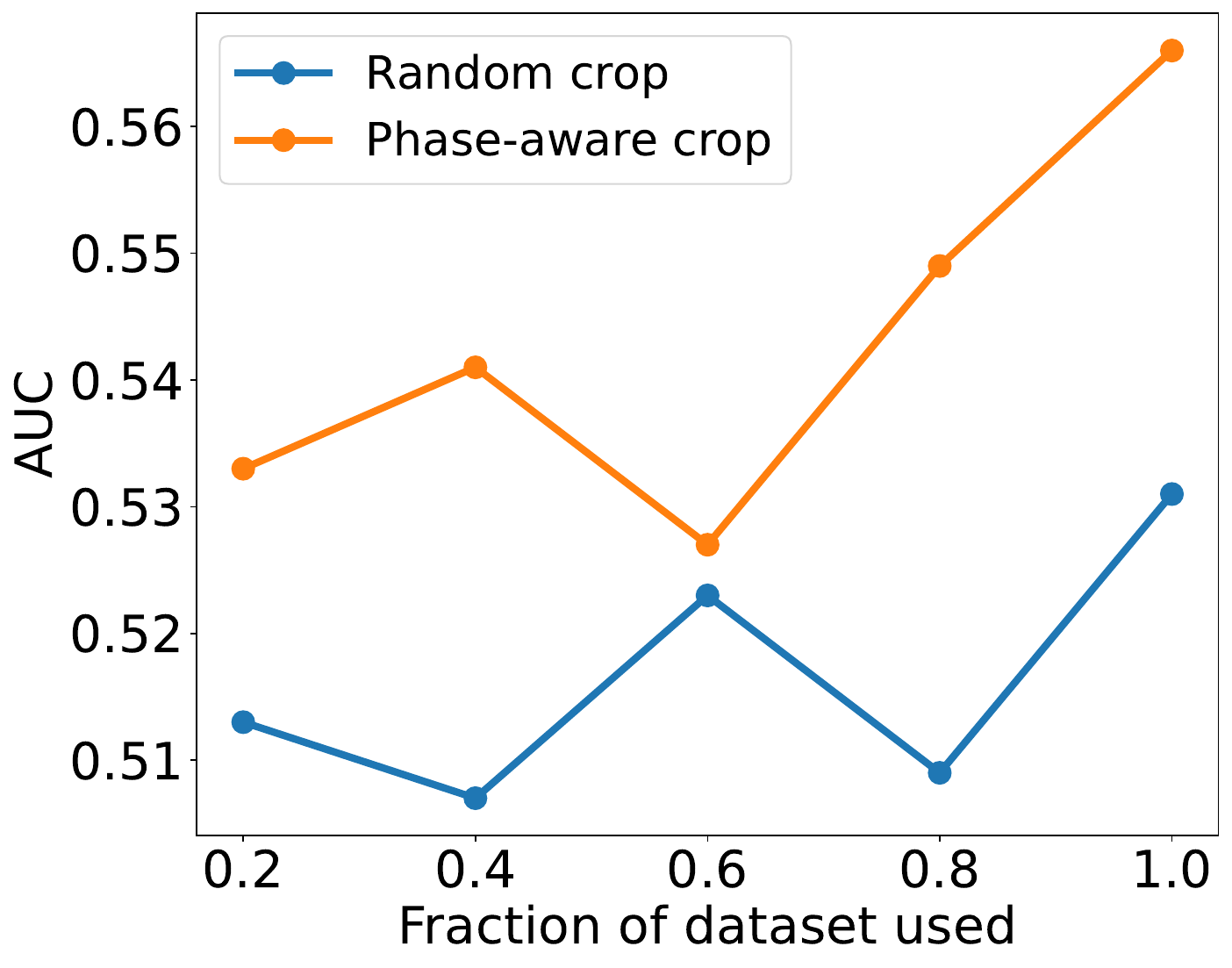}}~%
    \subfloat[\emph{COPD-DE}]{\includegraphics[width=.24\textwidth]{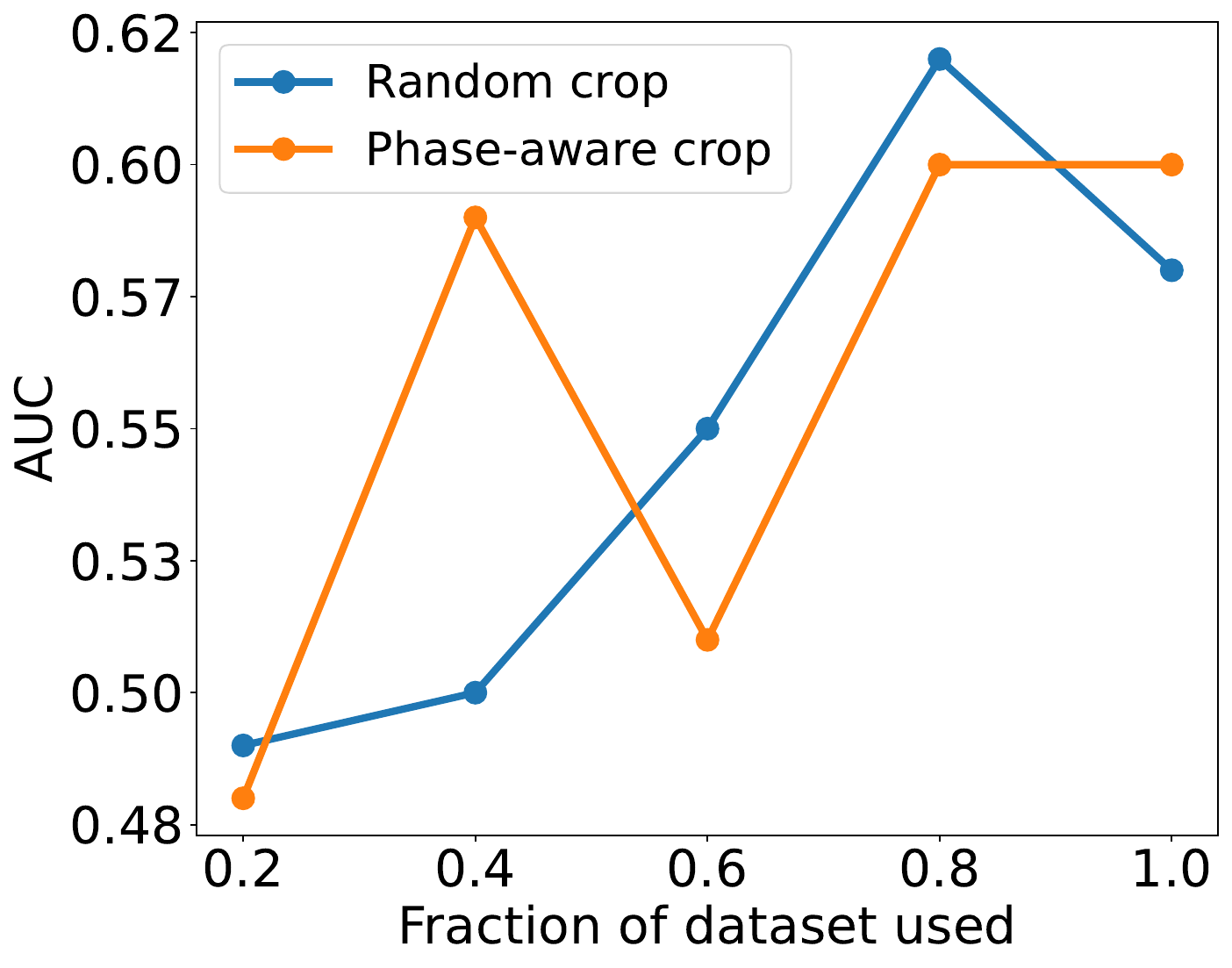}}
    
    \subfloat[\emph{COUGHVID} Gender]{\includegraphics[width=.24\textwidth]{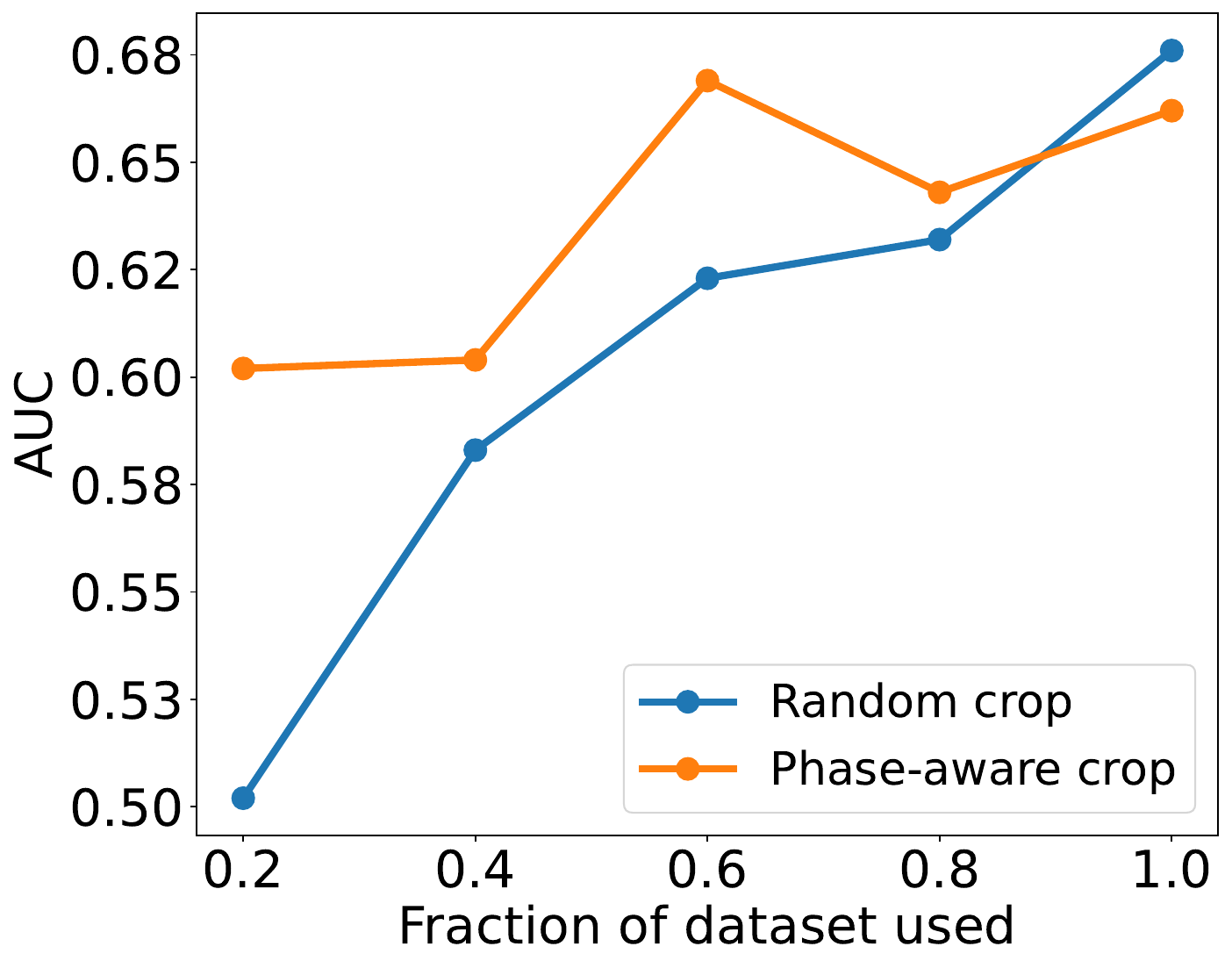}}~%
    \subfloat[\emph{Coswara} Smoker]{\includegraphics[width=.24\textwidth]{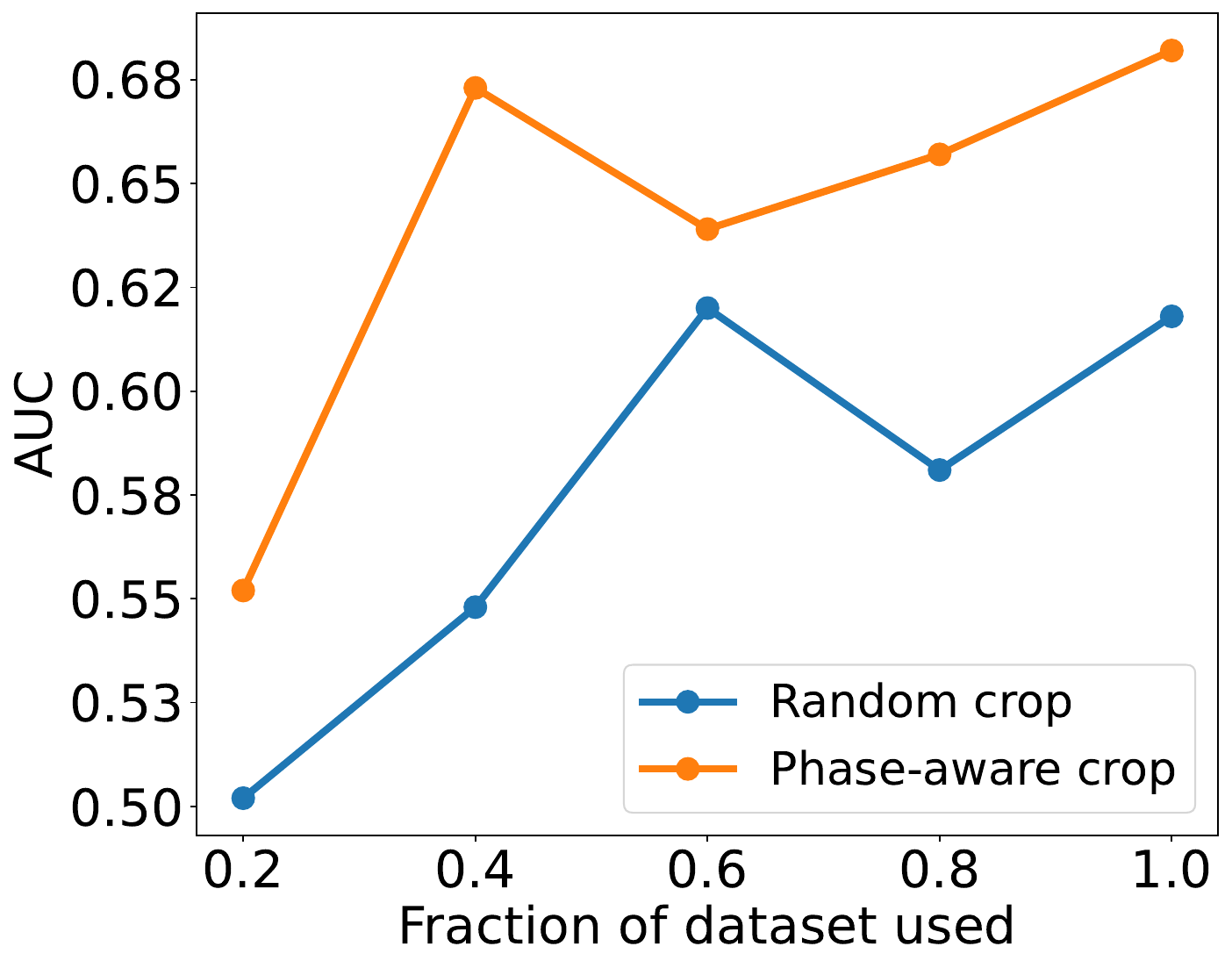}}~%
    \subfloat[\emph{Coswara} Gender]{\includegraphics[width=.24\textwidth]{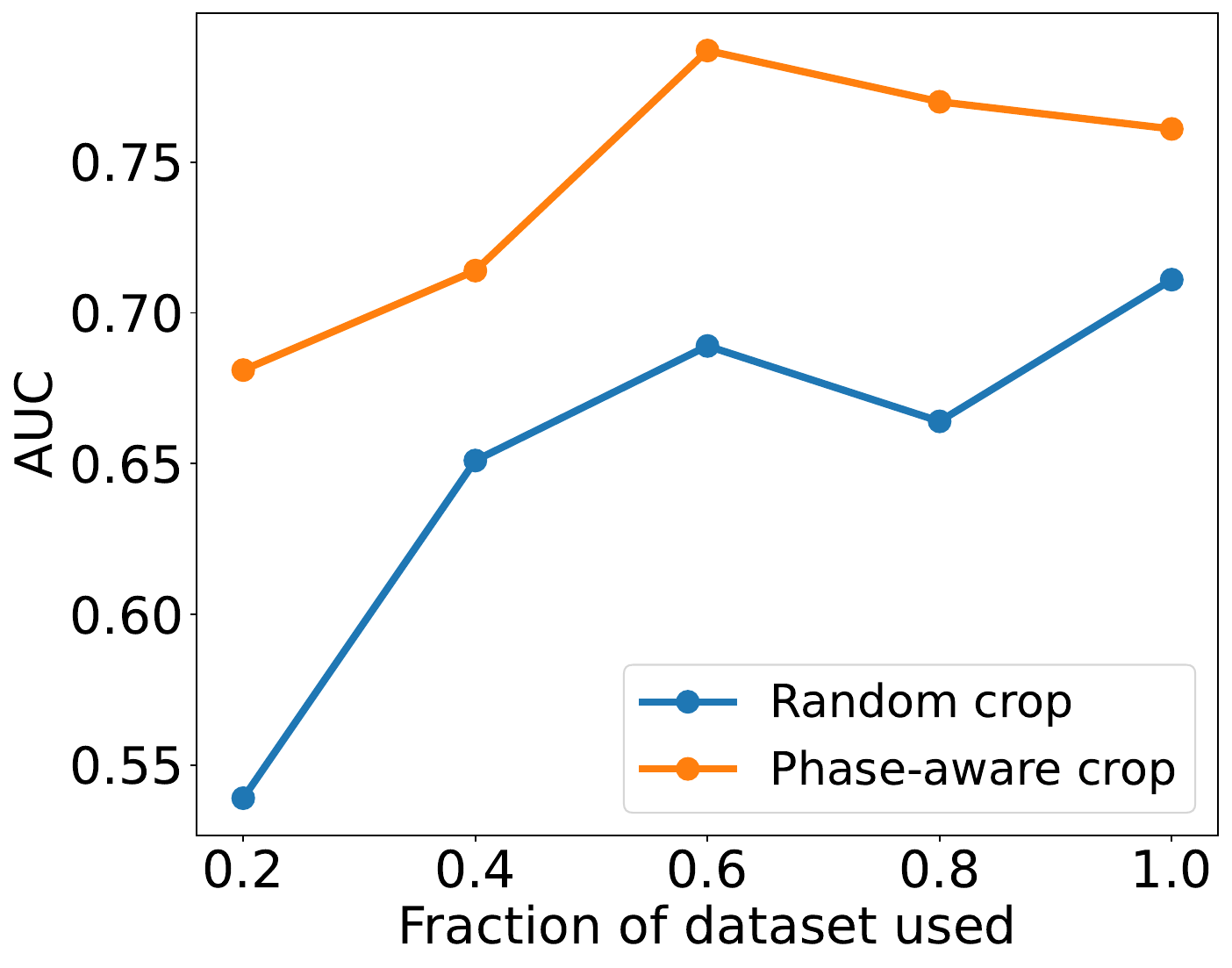}}
    
    \caption{Comparison of the two pre-training tasks, random crop (\emph{OPERA-CE-Cough}) and phase-aware crop (\emph{CoughPhase-CLR}), using different fractions of the dataset for pre-training.}
    \label{fig:data_efficiency}
\end{figure*}

Results can be seen in \cref{fig:data_efficiency}.
Generally, both models showed an increase in downstream performance as the size of the pre-training data increased.
Importantly, \emph{CoughPhase-CLR} outperformed \emph{OPERA-CE-Cough} in 4/5 tasks irrespective of the size of the pre-training data.
This suggests that the phase-aware contrastive task shows a better utilization of the pre-training data overall and is expected to scale better as more data is used for pre-training.

\subsection{Saliency Map Analysis}
\begin{figure*}[t]
    \centering
    \subfloat[\emph{OPERA-CE-Cough} Exacerbation]{\includegraphics[width=.31\textwidth]{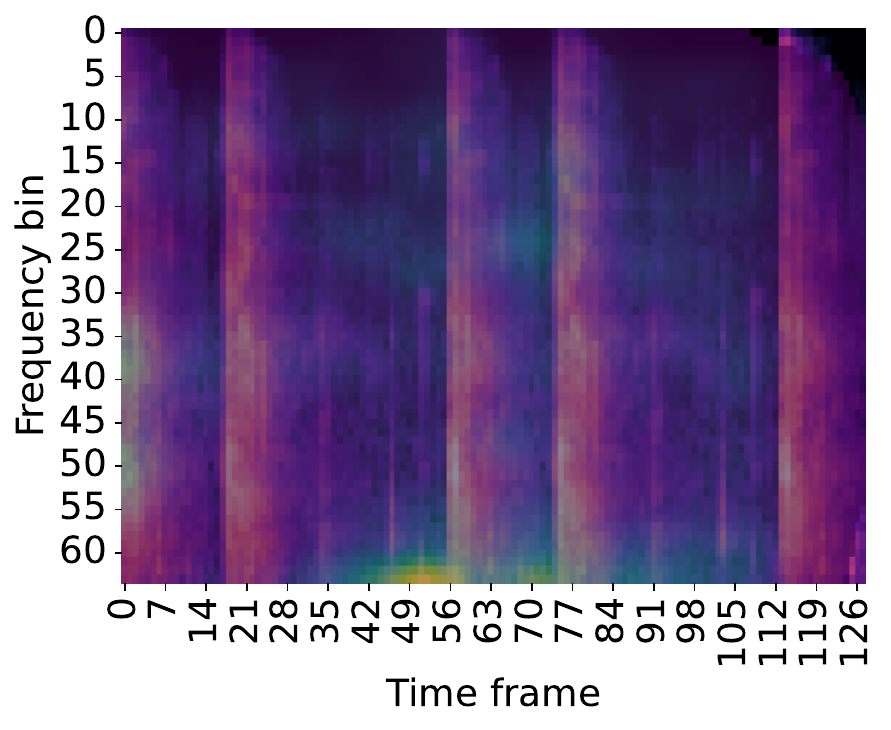}}~%
    \subfloat[CoughPhase-CLR Exacerbation]{\includegraphics[width=.31\textwidth]{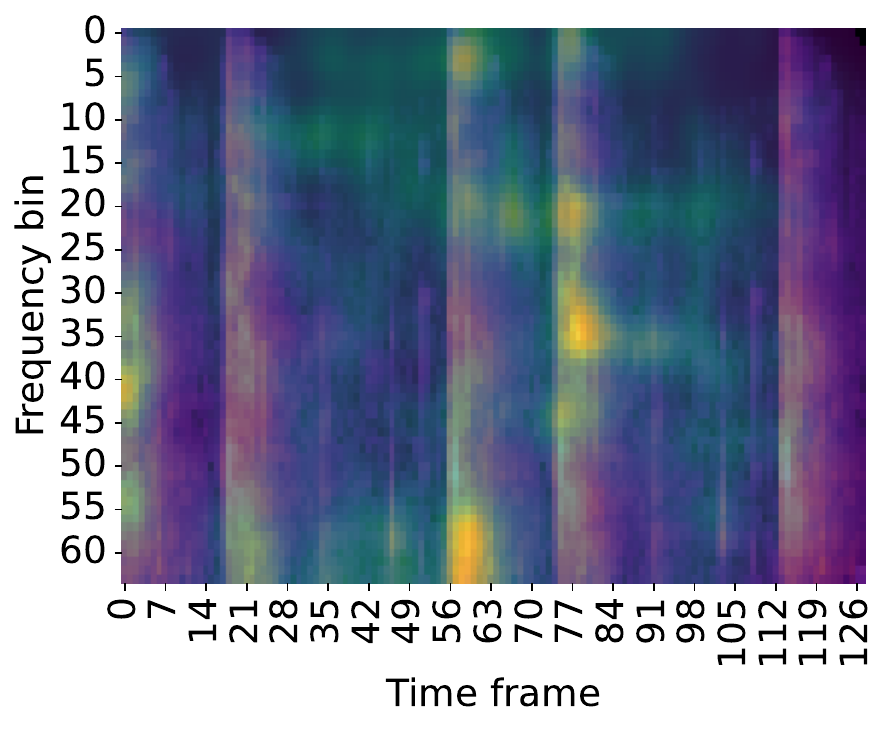}}~%
    \subfloat[OPERA-CE Exacerbation]{\includegraphics[width=.31\textwidth]{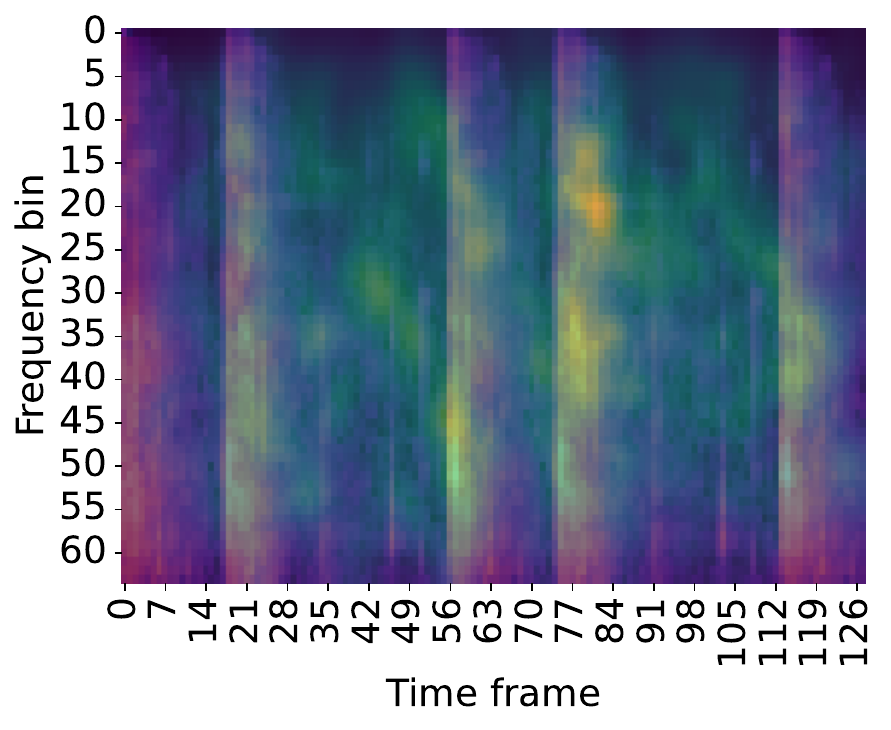}}

    \subfloat[\emph{OPERA-CE-Cough} Stable]{\includegraphics[width=.31\textwidth]{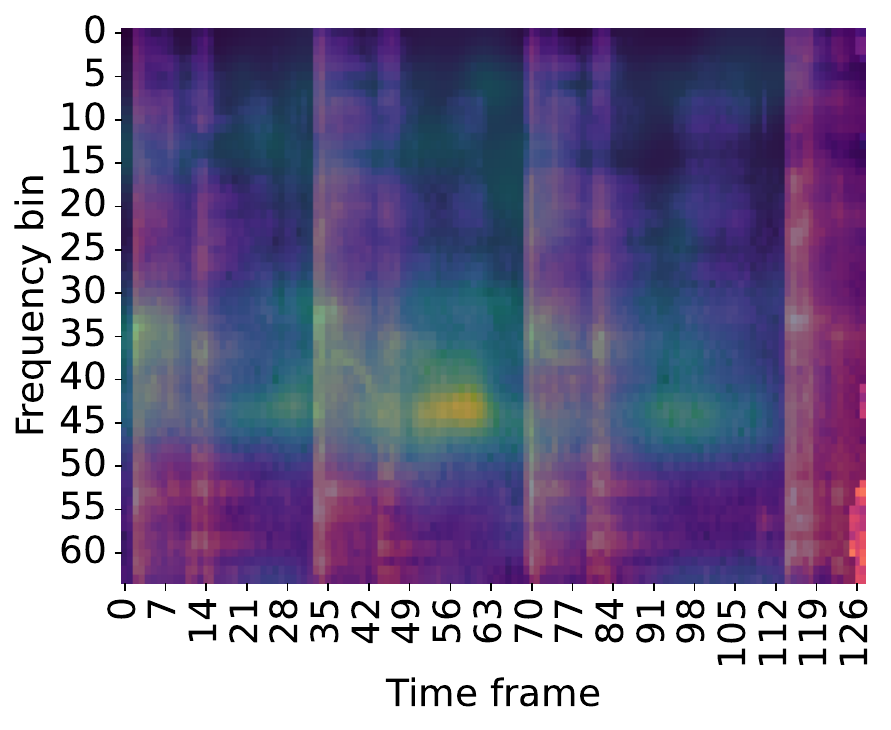}}~%
    \subfloat[\emph{CoughPhase-CLR} Stable]{\includegraphics[width=.31\textwidth]{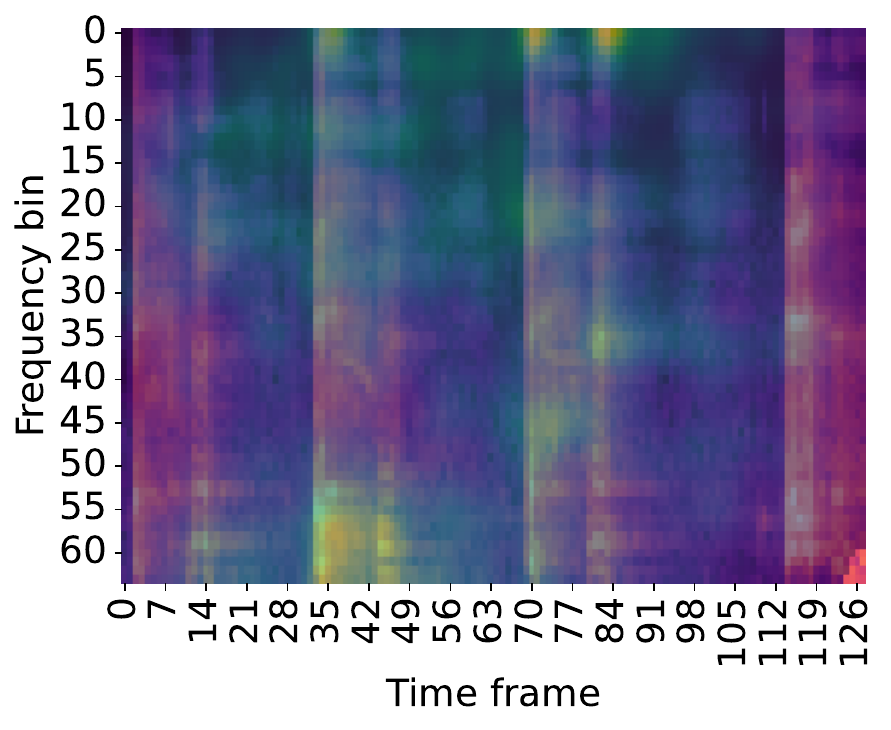}}~%
    \subfloat[OPERA-CE Stable]{\includegraphics[width=.31\textwidth]{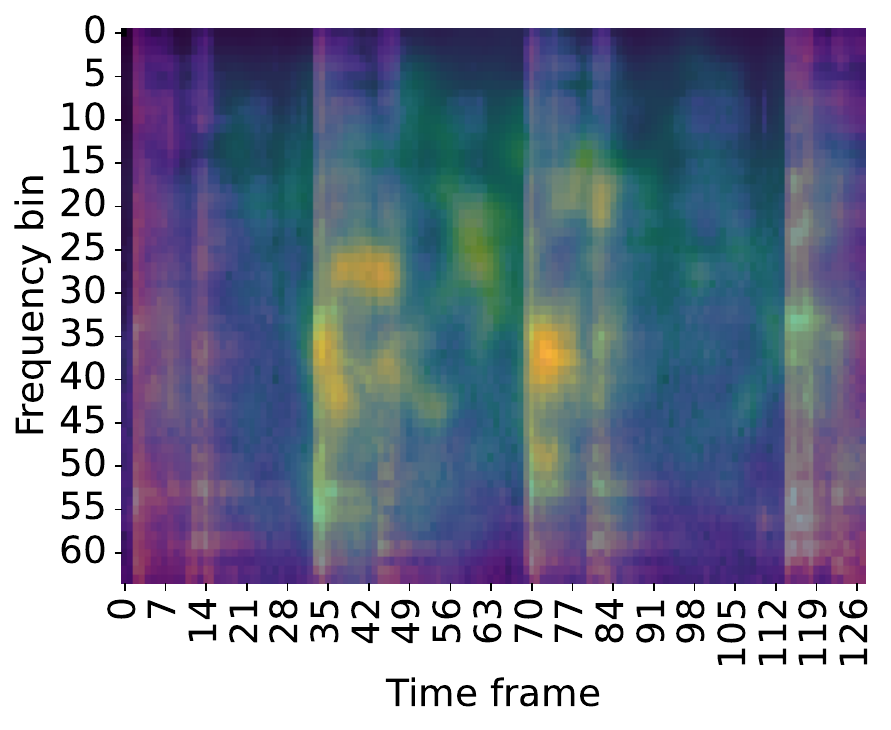}}
    \caption{Saliency maps illustrating model interpretability on cough spectrograms. The top row displays a cough from the exacerbation class, while the bottom row shows one from the stable class. Brighter areas correspond to input features with higher importance for the model's classification decision.}
    \label{fig:saliency_maps_combined}
\end{figure*}

An analysis of gradient-based saliency maps reveals that both the pre-training strategy, as well as the pre-training dataset influence how models attend to features on spectrograms. As shown with representative examples in \cref{fig:saliency_maps_combined}, models pre-trained on diverse respiratory sounds (\emph{OPERA-CE}) or with a cough-specific task (\emph{CoughPhase-CLR}) learn to focus on more meaningful time-frequency structures, compared to using a more generic task when pre-training only on coughs (\emph{OPERA-CE-Cough}).

Overall, the three models exhibit distinct spatial patterns in the time–frequency plane:
\emph{OPERA-CE} shows a rather diffuse attention, highlighting major parts of the coughs very consistently but also considering the context between coughs. This behaviour is likely inherited from its pre-training on diverse respiratory events with longer temporal structure such as breathing and lung sounds, encouraging reliance on broader context. This likely helps focus more consistently on coughs, leading to the superior and more robust classification performance.
\emph{OPERA-CE-Cough} demonstrates poor feature localization, focusing on areas with little acoustic energy related to the cough itself, such as the low-frequency bands or inter-cough regions. This suggests its pre-training strategy was least effective at identifying salient cough characteristics.
\emph{CoughPhase-CLR} shows a higher focus on the high energy parts of the cough compared to \emph{OPERA-CE-Cough}, demonstrating the success of the cough-specific pre-training task. It, however, also has the tendency to focus on high-frequency regions, which likely contribute little meaningful information and may hinder the model's performance.

In summary, while our phase-aware pre-training method successfully encourages the model to increase its focus on the cough onsets, the broader contextual awareness of \emph{OPERA-CE} appears more effective for the downstream classification task. Nevertheless, saliency maps are heuristic and do not establish causality, meaning that they should be interpreted cautiously.

\subsection{COPD Benchmark}
\label{sec:copd_classification_results}

This section presents more detailed results on the COPD classification task, with a summary of the results available in \cref{tab:uar-results}.
None of the tested models is able to obtain comparable performance to the state-of-the-art result of 84\% (achieved using speech features) \citep{COPDSpeech2Triantafyllopoulos}, indicating how challenging cough classification is.
The best performance of 57\% was achieved by \emph{OPERA-CT}, as well as \emph{CNN14} pre-trained on \emph{AudioSet}, and was followed closely by \emph{OPERA-CE} with 56\% UAR.

Using the \emph{eGeMAPS} features only achieved a UAR of 51\%, barely performing better than random chance. This suggests that using interpretable features does not work well for this task. Comparing this to the performance of the non-pre-trained deep neural networks, we see that \emph{CNN14} performed better than the handcrafted features with a UAR of 54\%. In contrast, \emph{EfficientNet-B0} (scratch) was only able to match its performance, also achieving 51\% UAR.
While \emph{CNN14}, pre-trained on \emph{AudioSet}, achieves comparatively good results, the models pre-trained solely on speech audio, \emph{wav2vec2.0} and \emph{HuBERT}, only achieved a UAR of 50\%, performing worse than all other tested models.

Models pre-trained on respiratory audio show promising results, but there is a notable inconsistency across different architectures and pre-training strategies. Between the \emph{OPERA} models, \emph{OPERA-CT} achieved the highest score in our evaluation with a UAR of 57\%, with \emph{OPERA-CE} close behind, achieving a UAR of 56\%. In contrast to this, the generatively pre-trained \emph{OPERA-GT} only performed at chance level with a UAR of 50\%, demonstrating that the pre-training task can be critical for success. \emph{HeAR} also showed poor performance, only achieving a UAR of 51\%.
Our proposed model \emph{CoughPhase-CLR} shows inferior performance compared to the contrastive pre-trained \emph{OPERA} models, only achieving a UAR of 53\%.

\begin{table}[t]
    \centering
    \caption{Comparison of COPD classification results. Results are given as UAR [\si{\percent}] with 95\% CIs and a chance level of 50\%.}
    \label{tab:uar-results}
    \begin{tabular}{lc}
        \toprule
        \textbf{Method}                                & \textbf{UAR [\si{\percent}]} \\
        \midrule
        \multicolumn{2}{l}{\textbf{1. No pre-training}}                               \\
        \cmidrule(r){1-2}
        \emph{eGeMAPS} \citep{eGeMAPS}                         & $51 [50-52]$                 \\
        \emph{CNN14} (scratch) \citep{PANNs}                   & $54 [52-57]$                 \\
        \emph{EfficientNet-B0} (scratch) \citep{EfficientNet}  & $51 [48-54]$                 \\
        \midrule
        \multicolumn{2}{l}{\textbf{2. Pre-trained on image data}}                     \\
        \cmidrule(r){1-2}
        \emph{EfficientNet-B0} (\emph{ImageNet}) \citep{EfficientNet} & $53 [50-56]$                 \\
        \emph{VGG-16-BN} \citep{VGG}                           & $51 [47-56]$                 \\
        \midrule
        \multicolumn{2}{l}{\textbf{3. Pre-trained on general audio data}}             \\
        \cmidrule(r){1-2}
        \emph{CNN14} (\emph{AudioSet}) \citep{PANNs}                  & $57 [52-61]$                 \\
        \emph{wav2vec2.0} \citep{wav2vec2.0}                   & $50 [48-52]$                 \\
        \emph{HuBERT} \citep{HuBERT}                           & $50 [48-51]$                 \\
        \midrule
        \multicolumn{2}{l}{\textbf{4. Pre-trained on respiratory audio}}              \\
        \cmidrule(r){1-2}
        \emph{HeAR} \citep{HeAR}                               & $51 [50-53]$                 \\
        \emph{OPERA-CE} \citep{OPERA}                          & $56 [54-57]$                 \\
        \emph{OPERA-CT} \citep{OPERA}                          & $57 [53-60]$                 \\
        \emph{OPERA-GT} \citep{OPERA}                          & $50 [47-54]$                 \\
        \midrule
        \multicolumn{2}{l}{\textbf{5. Pre-trained on cough audio}}                    \\
        \cmidrule(r){1-2}
        \emph{CoughPhase-CLR} (ours)                          & $53 [51-55]$                 \\
        \bottomrule
    \end{tabular}
\end{table}

\begin{figure*}[t]
    \centering
    \includegraphics[width=0.95\textwidth]{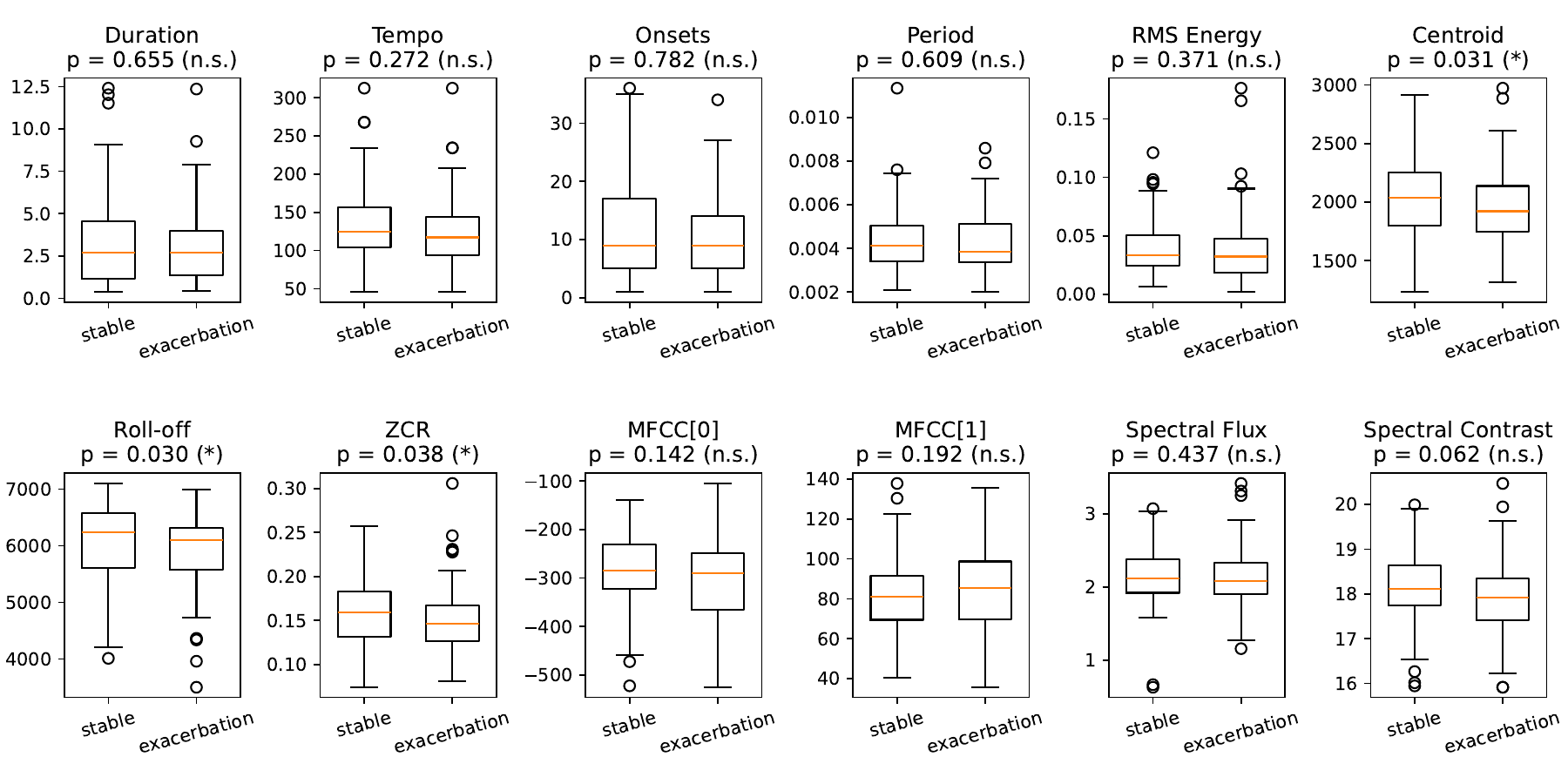}
    \caption{Box plots of twelve acoustic features extracted from cough
    recordings, compared between the stable and exacerbation states.
    Each box shows the median and interquartile range (IQR), with whiskers
    extending to 1.5$\times$IQR and outliers plotted individually. The
    abbreviated features are root-mean-square (RMS) energy, zero-crossing
    rate (ZCR), and the first two mel-frequency cepstral coefficients
    (MFCC[0], MFCC[1]). Above each panel, the p-value from a two-sided
    Mann--Whitney U test is reported; an asterisk ($*$) marks a
    statistically significant difference at $\alpha = 0.05$ and ``n.s.''
    denotes a non-significant result.}    
    \label{fig:data_analysis}
    \end{figure*}

To try to explain why the classification of COPD states using coughs proves so difficult, we performed an analysis of various interpretable acoustic features. Figure~\ref{fig:data_analysis} presents box plots of these features for the stable and exacerbation states, annotated with p-values from two-sided Mann–Whitney U tests. The distributions overlap substantially between the two states, with the majority of features showing no statistically significant difference at $\alpha = 0.05$.
This suggests that handcrafted acoustic features may be insufficient for this task.
While these features are not directly used as model inputs in our deep learning models, they are linked to the acoustic content of the input data.

In contrast to the findings for cough audio, the speech features analyzed in \citep{COPDSpeech2Triantafyllopoulos} showed a decrease in spectral flux, reduced loudness variance, and a lower bandwidth spread following an exacerbation.
This is partially explained by the fact that \emph{eGeMAPS} has been primarily designed for speech signals.
As a result, many of its features are not relevant for coughing; e.\,g. F0-derived features are all ill-defined for coughing, where, typically, no fundamental frequency is detected.
Nevertheless, many spectral features, such as spectral roll-off or flux, are indeed relevant for coughing, as they relate to the different stages of a coughing signal.
This indicates the inherent difficulty of classifying COPD states using handcrafted audio features, highlighting the need for more powerful models that can discover the most discriminative features by themselves.

\subsection{Data Efficiency of Pre-trained Models}

\paragraph{Performance with Limited Pre-training Data}
While \emph{CoughPhase-CLR} consistently outperforms \emph{OPERA-CE-Cough} when pre-trained on all data, this performance gap widens when pre-training on smaller data subsets, suggesting that the phase-aware contrastive task is more data-efficient than random cropping.
In contrast, a less constrained approach like random cropping might struggle when the training data is not diverse enough, leading the model to overfit on less relevant regions for cough classification.

\paragraph{Performance with Larger Pre-training Datasets}
On larger datasets, \emph{OPERA-CE-Cough} is sometimes able to close the performance gap to \emph{CoughPhase-CLR} or even overtake it.
A possible explanation for this could be that random cropping increases the diversity of observed windows across epochs, which can help the model learn more robust cough representations, resulting in better generalization. Additionally, random cropping in \emph{OPERA-CE-Cough} does not discard information like \emph{CoughPhase-CLR}, which removes the patients' inhalation and is not able to fully capture the transition between the first two cough phases in one view, as this is where the cough is segmented into the two views.
However, this trend cannot be observed for all tasks, since in T1, the gap widens as the pre-training dataset grows, suggesting that this benefit is task-dependent.

\section{Discussion}
\label{sec:discussion}

Our experimental results provide several key insights into the differences of pre-training strategies.
We showed that a cough-specific, phase-aware pre-training strategy (\emph{CoughPhase-CLR}) is more beneficial than standard random cropping (\emph{OPERA-CE-Cough}) when pre-training on cough audio. This outcome aligns with the results found in \citet{GoodContrastiveTask}, which showed that optimal views in contrastive learning are task-dependent and preserve label-relevant content while removing irrelevant nuisance.
The improved performance of \emph{CoughPhase-CLR} suggests that the quality of the contrastive views is ultimately more important than their temporal diversity, achieved through random cropping, on our set of downstream tasks. By constructing the positive pairs around the main energy peaks within a cough event, \emph{CoughPhase-CLR} is able to produce more informative views more consistently, which allows the encoder to focus on the cough events instead of focusing on non-informative views or noise. \emph{OPERA-CE-Cough}, on the other hand, tends to generate less informative views that contain silence, or produces two overlapping views as positive pairs, making the contrastive task easier for the model.
While we were not able to achieve superior performance compared to \emph{OPERA-CE}, it is important to note that \emph{OPERA-CE} was trained on over 400 hours of audio while both \emph{CoughPhase-CLR} and \emph{OPERA-CE-Cough} were trained on under 40 hours of cough audio. This significant difference in pre-training dataset size likely contributes to the superior performance of \emph{OPERA-CE} across most tasks. The increased amount of training data enables the model to learn more robust and generalizable representations, which is reflected in its higher AUROC scores.

Another primary finding of this work is the inherent difficulty of classifying COPD states using cough audio. While the top-performing models were able to consistently achieve results higher than chance level, they failed to approach the state-of-the-art results previously established on the same dataset using speech analysis~\citep{COPDSpeech2Triantafyllopoulos}. This gap suggests that cough sounds are less informative than speech signals in distinguishing between stable and exacerbation states in COPD patients and illustrates the necessity of developing more powerful foundation models for coughing.

Beyond that, we evaluated a wide range of pre-training strategies for \emph{EfficientNet-B0}. A primary finding is that all forms of pre-training achieved superior performance compared to training the model from scratch. \emph{OPERA-CE} \citep{OPERA} achieved the highest performance, outperforming both \emph{EfficientNet-B0} (\emph{ImageNet}) and \emph{CoughPhase-CLR} which achieved the same performance.
The lower performance of the \emph{ImageNet}-based pre-training suggests that features learnt from natural images are less transferable to spectrograms compared to also pre-training on spectrograms.
\emph{OPERA-CE} was, however, able to outperformed \emph{CoughPhase-CLR}, indicating that pre-training beyond only cough audio by also including a variety of other respiratory sounds boosts performance for COPD state classification.
Comparing the performance between the \emph{OPERA} models, it becomes apparent that the contrastive pre-trained \emph{OPERA-CT} and \emph{OPERA-CE} achieve notably higher performance than the generatively pre-trained \emph{OPERA-GT}. This aligns with the findings in \citep{OPERA,DiverseDatasetBetterThanOPERA,COPDCoughSAMOPERA}, which state that \emph{OPERA-GT} performs worse on classification tasks, but beats the contrastive pre-trained \emph{OPERA} models on the regression task.
While the best-performing models were pre-trained, it did not always guarantee superior performance. Pre-trained models like \emph{VGG-16-BN}, \emph{wav2vec2.0}, and \emph{HuBERT} showed poor performance and were outperformed by \emph{CNN14} (scratch).

Nevertheless, our work features some limitations which can be improved in the future:
The size of our COPD-DE is relatively small; while this reflects the challenges of collecting real-world, clinically-validated audio data, it nevertheless can be improved by a more concerted data collection effort.
Our computational constraints also prevented us from conducting a large-scale comparison of hyperparameters; our recent work demonstrates the significance of that for audio benchmarks~\citep{Triantafyllopoulos26-C1Y}.
However, for most experiments, we chose the original hyperparameter set from \emph{OPERA-CE}; if anything, this reflects a positive bias towards our baseline and further optimization might benefit our proposed model instead.
A more in-depth comparison of the most promising foundation models for coughing can be conducted by a larger consortium with more computational resources.
Finally, we were unable to test our models under the more realistic scenario of multiple diseases and multiple comorbidities.
This limitation plagues most current works on diagnosing respiratory disorders, as very few datasets feature multiple, clinically-validated diagnoses.
This is something that the community should pay particular attention to.

To allow for real-world clinical applications, future work must improve the robustness of cough classification models. This requires further expanding the COPD dataset, either through crowdsourcing~\citep{sankey2025detecting}, as done for the COVID-19 datasets \citep{Coswara,Coughvid,UK_COVID19}, or through clinical studies, to improve the model's ability to generalize to new data.
Furthermore, the models should be evaluated on external datasets from different patient populations and recording conditions. Such testing is essential to test the model's real-world effectiveness and to analyze any biases learnt from the dataset.
Finally, a particularly promising approach is to combine multiple signals (connected speech, coughing, breathing, sustained vowels, or others) to improve the classification performance by leveraging all data available for each patient.

\section{Conclusion}
\label{sec:conclusion}
In this work, we presented \emph{CoughPhase-CLR}, a novel self-supervised learning framework that leverages the distinct acoustic phases of cough events to learn robust representations. 
Our results offer two distinct takeaways regarding representation learning. First, data volume remains a dominant factor, as the respiratory-pre-trained \emph{OPERA-CE}, trained on over 400 hours of audio, generally outperformed our model, which was trained on less than 40 hours. However, when controlling for data size, \emph{CoughPhase-CLR} demonstrated superior data efficiency and performance compared to standard random-cropping approaches (\emph{OPERA-CE-Cough}).
An analysis of salience maps further revealed that our approach successfully guided the model to focus on high-energy cough onsets rather than background noise.
The classification of COPD states, however, remained challenging, with the best-performing models OPERA-CT and CNN14 pre-trained on AudioSet, reaching a UAR of 57\%, compared to a performance of 84\% using speech. 
Ultimately, while \emph{CoughPhase-CLR} establishes a more effective method for learning from cough signals, future work should focus on (i) multimodal fusion of coughs with speech, sustained vowels, and clinical test results, (ii) expanding respiratory datasets using diverse audio recordings to improve generalization across different patient populations and recording environments, and (iii) further exploring the phase-aware pre-training strategy by expanding the pre-training dataset and using more advanced architectures.

\section{\refname}
\printbibliography[heading=none]



\end{document}